\documentclass[12pt]{iopart}

\usepackage{iopams}
\usepackage{amsfonts,bbm}
\usepackage[mathscr]{eucal}
\usepackage{epsfig}

\newcommand{\aver}[1]{\langle #1 \rangle}

\newcommand{\ket}[1]{\left|{#1}\right\rangle}
\newcommand{\bra}[1]{\left\langle{#1}\right|}

\newcommand{\ketbrad}[1]{\left|{#1}\rangle\!\langle{#1}\right|}

\newcommand{\be}{\begin{equation}}
\newcommand{\ee}{\end{equation}}

\newcommand{\openone}{\mathbbm{1}}
\newcommand{\eqref}[1]{(\ref{#1})}
\newcommand{\binom}[2]{
\left( \begin{array}{c}
#1 \\
#2
\end{array}
\right)}

\begin{document}
\title{Beating noise with abstention in state estimation}
\author{Bernat Gendra, Elio Ronco-Bonvehi, John Calsamiglia, Ramon Mu\~{n}oz-Tapia, and Emilio Bagan}
\address{F\'{i}sica Te\`{o}rica: Informaci\'{o} i Fen\`{o}mens Qu\`antics, Universitat Aut\`{o}noma de Barcelona, 08193 Bellaterra (Barcelona), Spain}

\date{\today}
\begin{abstract}
We address the problem of estimating pure qubit states with non-ideal (noisy) measurements in the multiple-copy scenario, where the data consists of a number $N$ of identically prepared qubits. We show that the average fidelity~of the estimates can increase significantly if the estimation protocol allows for inconclusive answers, or abstentions. We present the optimal such protocol and 
compute its fidelity for a given probability of abstention. The improvement over standard estimation, without abstention, can be viewed as an effective noise reduction. 
These and other results are exemplified for small values of~$N$. For asymptotically large $N$, we derive
analytical expressions of the fidelity and the probability of abstention, and show that for a fixed fidelity gain  the latter decreases with $N$ at an exponential rate given by a Kulback-Leibler (relative) entropy.
As a byproduct, we obtain an asymptotic expression in terms of this very~entropy of the probability that a system of $N$ qubits, all prepared in the same state, has a given total angular momentum. 
We also discuss an extreme situation where noise increases with $N$ and where estimation with abstention provides a most significant improvement as compared to the standard approach. 
\end{abstract}


\maketitle

\section{Introduction}
Knowing the state of a system is a key task in quantum information processing. An unknown quantum state can only be unveiled by means of measurements.
These, however, provide only partial knowledge about the system and, furthermore, this information gain comes always at the expense of destroying the state.   Only when a
reasonably large number $N$ of identically prepared  copies of the system is available an accurate estimation of the state is possible.   For a given $N$, the aim is 
then to find the measurement protocol that yields the  best estimate of the input state.  

The standard estimation optimization problem is suited for a situation where, say, an experimentalist is confronted with an unknown state of a system of which she is asked to
provide an estimate, based of course on the results of a measurement of her choice. A quantitative assessment of her performance is usually given by the expected value of 
the fidelity (or some other distinguishability measure) between the unknown input and her guess (see below).
Hence, it is implicitly assumed that the experimentalist is obliged to provide such guess regardless of the measurement outcome she obtains.
In this context, many results have been obtained over the last years  in a large variety of scenarios~\cite{holevo,paris-rehacek,massar-popescu,hradil,derka-buzek,bruss,fischer,bagan-local,bagan-mixed,bagan-optimal,blume,steinberg,gross}.  

Here, we will study a variation of this setting suited for a situation where the experimentalist is allowed to decide whether to provide a guess or abstain from doing~so. Of course, this
decision cannot be based on the actual state of the system (which is unknown  by definition) but rather on the result of a measurement. This relaxation of the original setting is very useful because
it enables the experimentalist to post-select her measurement outcomes in order to provide a more accurate guess. That is, the possibility of abstaining enables her to discard
instances where the measurement outcome turns out not to be informative enough. We will find that abstention can provide an important advantage,
specially in noisy scenarios. The problem of  `state estimation with abstention' \footnote{We thank G.~Chiribella for suggesting this word.}  is specially relevant in
situations where the experimentalist can afford to re-run the experiment, i.e. she can easily prepare a new instance of the problem, or
where she prioritizes having high quality estimates. 

Post-selection is a widely used tool in quantum information, particularly in experimental scenarios, where one has special demands or constrains.
A form of abstention has been already explored  in state discrimination~\cite{bergourev}, another important quantum statistical inference primitive.  Discrimination aims at identifying in which, out of a set of known quantum states, a system has been prepared. Two fundamental approaches are usually considered: the so-called `minimum-error', where  the experimentalist always has to provide a conclusive answer, at the expense, of course, of being wrong with certain probability~\cite{helstrom}, and `unambiguous discrimination', where no errors are permitted, but instead an inconclusive answer (abstention) may be given with some probability~\cite{unambiguous}.  
By varying the allowed rate of inconclusive answers~\cite{chefles2,zhang,fiurasek,eldar}, we may go from one approach to the other~\cite{hayashi,sugimoto,bagan-q}. The possibility of abstaining has been studied in \cite{fiurasek2} for phase estimation, and in~\cite{in-preparation-1}  for direction estimation with arbitrary pure input signals. In both cases the results show a significant improvement over the standard  (without abstention) approach.

In this work we consider the optimal estimation of a completely unknown pure qubit state when $N$ copies of it are available for measurement  and when certain   
rate (probability and rate will be used interchangeably throughout  the paper) of abstention~$Q$ is permitted. We will show that in an ideal noise-free scenario, abstention does not improve the estimation accuracy. However, it does in a realistic noisy scenario, as we claimed above. Here we will consider a simplified model where noisy measurements will be replaced by local depolarizing channels followed by ideal measurements. 

The paper is organised as follows. In the next section we consider estimation without abstention. More precisely, we obtain the protocol that gives the best estimate of the state of a qubit  based upon non-ideal measurements on $N$
independent and identically prepared systems.
In Section~\ref{sec:abstention}, estimation with abstention is introduced, and the optimal protocol for a fixed value of the abstention rate $Q$ is obtained. We study the asymptotic
regime of large $N$ and derive the corresponding maximum fidelity and probability of abstention.  
As an example, we also consider an scenario where abstention gives a drastic improvement. This is the case when noise increases with~$N$ in such a way that the fidelity of the estimation approaches a finite value 
less than~one as~$N$ becomes large. 
We close  the paper with some brief conclusions and present an outlook 
for future work.

\section{No abstention}\label{sec:no-abstention}
Let us consider $N$ copies of Êa completely unknown pure qubit Êstate $\ket{\vec{n}}$ (throughout  the paper $\vec n$ will denote a unit Bloch vector) that we wish to estimate by performing a realistic, and therefore noisy, quantum measurement.
We model it as an ideal measurement preceded by the single-qubit depolarizing channel acting on every copy:

\begin{equation}\label{noise}
\mathcal{E}(\rho)=(1-\eta) \rho +\frac{\eta}{3}(\sigma_{x}\rho\sigma_{x}+\sigma_{y}\rho\sigma_{y}+\sigma_{z}\rho\sigma_{z}),
\end{equation}
where with probability $1-\eta$ no error occurs, while with probability $\eta$
the state is affected by
either a bit-flip, a phase-flip, or both. This error probability $\eta$ is assumed to be known by the experimentalist, therefore, for the purpose of analyzing the effects of noise in the estimation process, we will transfer its effect to the input states and optimize the estimation protocol over ideal measurements.
Hence, we will consider input states of the form
\begin{equation}\label{rho}
\rho(\vec{n})=r \ketbrad{\vec{n}} +(1-r)\frac{\openone}{2}=\frac{\openone+r\, \vec{n}\cdot \vec{\sigma}}{2},
\end{equation}
with $r=1-(4/3) \eta$. In words, we will assume that  the input states either do not change with probability $r$ or they become completely randomized with probability~\mbox{$1-r=(4/3) \eta$}.
The original problem is thus equivalent to
the estimation of a pure state~$|\vec n\rangle$  (or of a uniformly distributed Boch vector
 $\vec{n}$)  based upon the outcomes of an appropriate ideal measurement on $N$ copies of the mixed state $\rho(\vec n)$ in~Eq.~\eqref{rho}, i.e., on the state~$\rho(\vec{n})^{\otimes N}=\tau(\vec{n})$.

For each measurement outcome $\chi$ an estimate $|\vec n_\chi\rangle$ is provided according to some guessing rule $\chi\to |\vec n_\chi\rangle$. 
We choose to quantify the quality of the estimate by means of the squared overlap 
\begin{equation}
f(\vec{n},\vec{n}_\chi)= |\bra{\vec{n}} \vec{n}_\chi \rangle|^2,
\end{equation}
also known as the fidelity. 
The overall quality of the estimation protocol is then given by the average fidelity
\begin{equation}\label{FideSin}
F=\sum_\chi \int dn f(\vec{n},\vec{n}_\chi)\, p(\chi|\vec{n}),
\end{equation}
where $dn=\sin\theta d\theta\, d\phi /(4\pi) $ is the uniform probability distribution on the two-sphere and $p(\chi|\vec{n})$ is the conditional
probability of obtaining the outcome $\chi$ if the input state is~$\tau(\vec{n})$. This probability is given by the Born rule $p(\chi|\vec{n})=\tr[\Pi_\chi \tau(\vec{n})]$, where $\Pi_\chi\geq 0$ are the elements of a Positive Operator Valued Measure (POVM). They satisfy the completeness relation $\sum_\chi \Pi_\chi=\openone$, where~$\openone$ denotes the identity operator in the space spanned by the input states $\{\tau(\vec{n})\}$.
The index $\chi$ may be discrete, continuous or both. 
 
A protocol (i.e., a measurement $\{\Pi_\chi\}$ and a guessing rule $\chi\to \ket{\vec{n}_\chi}$) is said to be optimal if it maximizes $F$.
 For pure states, $r=1$, the maximum fidelity is well-known~\cite{massar-popescu}:
\begin{equation}\label{massar-popescu}
F=\frac{N+1}{N+2}=1-{1\over N}+\mathscr{O}(N^{-2}).
\end{equation}
It is also known that the (continuous) covariant POVM
\begin{equation}\label{povm-pure}
\Pi(\vec{s})=(2J+1) U(\vec{s})\ketbrad{J\; J} U^{\dag}(\vec{s})
\end{equation}
(with the obvious guessing rule $\Pi(\vec s)\to |\vec s\rangle$) is optimal. In~(\ref{povm-pure}), we use the standard notation, where $\{|jm\rangle\}_{m=-j}^j$ is the eigenbasis of the total angular momentum operators~$J^2$ and $J_z$. 
We~denote by $U(\vec{s})=[u(\vec s)]^{\otimes N}$, $u(\vec s)\in\mbox{SU(2)}$, (the unitary representation of) the rotation that maps the unit (Bloch) vector $\hat z$ into $\vec s$ [thus~$u(\vec s)|\mbox{\footnotesize ${1\over2}{1\over2}$}\rangle=|\vec s\rangle $],
and we have also introduced the definition~$J\equiv N/2$. Note that 
the POVM $\{\Pi(\vec s)\}$ 
acts on the symmetric
subspace of largest total angular momentum~$J$, 
of dimention~$2J+1=N+1$.  
In terms of~$J$, \eqref{massar-popescu} can also be written as
\begin{equation}\label{f-pure}
F=\frac{1}{2}\left( 1+\frac{J}{J+1}\right)\equiv \frac{1}{2}\left( 1+\Delta_J\right).
\end{equation} 

Mixed states span a much larger Hilbert space and the computation becomes more involved.  It greatly simplifies in the total angular momentum basis, where
the input state $\tau(\vec{n})$ is block-diagonal~\cite{bagan-mixed}. We have
\begin{equation}\label{tau-n}
\tau(\vec{n})=\sum_{j=j_\mathrm{min}}^{J} \sum_{\alpha=1}^{n_j} p_{j\alpha} \tau_{j\alpha} (\vec{n}),
\end{equation}
where $ \tau_{j\alpha}(\vec n)$
is the  normalized mixed state  
\begin{equation}\label{tau-j-alpha}
\tau_{j\alpha}(\vec{n})=\frac{1}{Z_j} \sum_{m=-j}^{j} R^m\; U(\vec n) \ket{jm;\alpha}\bra{jm;\alpha} U^\dagger(\vec n),
\end{equation}
with the definitions:
\begin{equation}\label{Z}
Z_j=\sum_{m=-j}^{j}R^m=\frac{R^{j+1}-R^{-j}}{R-1},\quad R=\frac{1+r}{1-r}>1.
\end{equation}
The additional  index $\alpha$, where $\alpha=1,2,\dots,n_j$, labels the various occurrences of the irreducible representation of total angular momentum $j$. The multiplicity $n_j$ is given by~\cite{cirac,bagan-mixed}
 \begin{eqnarray}\label{nj}
 n_j &=& \binom{2J}{J-j} -\binom{2J}{J-j-1}\nonumber \\[.2em]
       &=&\binom{2J}{J-j} \frac{2j +1}{J+j+1}.
 \end{eqnarray}
 In the sum~(\ref{tau-j-alpha}), $j$ runs from 
 $j_\mathrm{min}=0$ ($j_\mathrm{min}=1/2$) for
 $N$ even (odd) to the maximum total angular momentum~$J$,
in contrast to the pure state case where only the maximum value 
$J$ appears.
The numbers $p_{j\alpha}>0$  are the probabilities that the state $\tau(\vec{n})$ has quantum numbers  $j$  and $\alpha$, i.e.,
$p_{j\alpha} =\tr [\openone_{j\alpha} \tau(\vec{n})]$, where $\openone_{j\alpha}=\sum_{m=-j}^j|jm;\alpha\rangle\langle jm;\alpha|$ is the projector onto the corresponding eigenspace. 
The projector onto the whole subspace of total angular momentum~$j$ is then
\begin{equation}\label{completeness}
\openone_j=\bigoplus_{\alpha=1}^{n_j} \openone_{j\alpha} .
\end{equation}
Since the input state is permutation invariant (under the interchange of the individual qubits) representations with the same
$j$ are just mere repetitions of the same representation, they contribute a multiplicative factor 
of~$n_j$ to the fidelity through the marginal probability 
$p_j=\sum_\alpha p_{j\alpha}$, which reads
\begin{eqnarray}
p_j&=&\left( \frac{1-r^2}{4}\right)^{J}n_j Z_j =\nonumber\\
&=& \left( \frac{1-r^2}{4}\right)^{J}\binom{2J}{J-j} \frac{2j +1}{J+j+1} 
\frac{R^{j+1}-R^{-j}}{R-1}.\label{p-j}
\end{eqnarray}
One can easily check that $\sum_j p_j=1$, as it should be. 

Because of the block diagonal form of the input states, an obvious optimal measurement  consists of a direct sum of  covariant POVMs,
 \begin{equation}\label{repeated}
 \Pi(\vec{s})=\bigoplus_{j=j_{\mathrm{min}} }^{J}\bigoplus_{\alpha=1}^{n_j}\Pi_{j\alpha}(\vec{s}),
 \end{equation}
 where each of them is a straightforward 
generalization of Eq.\eqref{povm-pure}:
 \begin{equation}\label{Pi-j-alpha}
 \Pi_{j\alpha}(\vec{s})=(2j+1)\, U(\vec{s})\ketbrad{j\; j;\alpha} U^{\dag}(\vec{s}).
 \end{equation}
 One can easily check that the completeness condition $\int ds \,\Pi(\vec{s})=\openone$ holds.
The total fidelity then is
  \begin{equation}\label{fid-r}
 F=\frac{1}{2}\left( 1+ \sum_{j=j_{\rm min}}^J p_j \Delta_j \right) ,
  \end{equation}
 where \cite{holevo}
\begin{equation}\label{Delta-j}
\Delta_j=\frac{\aver{J_z}_j}{j+1}=\frac{\tr [J_z \,\tau_j(\hat{z})]}{j+1},
\end{equation}
with $\tau_j(\hat{z})$ being any one of the normalized states defined in Eq.~\eqref{tau-j-alpha},
(say, the one~with~$\alpha=1$). A straightforward calculation gives
\begin{equation}\label{m-bar}
\aver{J_z}_j=\frac{1}{Z_j} \sum_{m=-j}^{j} m R^m=j-\frac{1}{R-1}+\frac{2j+1}{R^{2j+1}-1}.
\end{equation}
Notice that for pure states, one has $R\to\infty$, and in turn $\aver{J_z}_J\to J$,  in agreement with~Eq.~\eqref{f-pure}.

As will be shown in the next section,  for asymptotically large $N$ the probability~$p_j$ peaks at 
 a value of $j\simeq r J$, which gives the dominant and subdominant  contributions to the sum in~\eqref{fid-r}. Up to order $1/N$, and discarding exponentially vanishing contributions [e.g., $\sim R^{-rJ}$], the asymptotic fidelity turns out to be
 \begin{equation}\label{fid-p}
 F=1- \frac{1}{Nr}\frac{r+1}{2 r}+\cdots \; .
 \end{equation}
This result is interesting on its own and, to the best of our knowledge, has not been presented before.
Note that for pure states ($r=1$) Eq.~\eqref{fid-p} agrees with the asymptotic expression of the fidelity in~Eq.~\eqref{massar-popescu}.

\section{Abstention}\label{sec:abstention}

In this section we  focus on estimation protocols where the experimentalist is allowed not to produce an answer, or abstain, if the outcome of the measurement she performed cannot provide a good enough estimate of the unknown state. Obviously, $F$ cannot decrease by excluding these abstentions from the average.  In noisy scenarios, such as that considered in this paper, $F$ actually increases, as will be shown below. Our aim is to quantify this gain and find the optimal protocol. In our approach, the probability of abstention, $Q$, is kept fixed, rather than unrestricted, since usually in practical situations one cannot afford discarding an unlimited amount of resources/state preparations.

\subsection{General framework}

To enable the possibility of abstaining, the POVM representing the measurement must include the abstention operator, which we denote by $\Pi_0$, in addition to the operators~$\{\Pi_\chi\}$, each of them associated to a specific estimate~$|\vec n_\chi\rangle$. Thus, the completeness relation reads
\begin{equation}
\sum_\chi \Pi_\chi +\Pi_0=\openone.
\end{equation}
The probability of abstention (abstention rate) and that of producing an estimate (acceptance rate) are then given respectively by
\begin{equation}
Q=\int dn\, \tr\left[\Pi_0\tau(\vec n)\right]\  \mbox{and}\ \bar Q=1-Q,
\label{def Q}
\end{equation}
and the mean fidelity defined in~(\ref{FideSin}) becomes now
\begin{equation}
F(Q)=\frac{1}{\bar Q}\sum_\chi\int dn\, f(\vec n,\vec n_\chi)\tr\left[\Pi_\chi \tau(\vec n)\right]
\label{F(Q)},
\end{equation}
where notice that the sum does not include the $\Pi_0$ operator and  $\bar Q$ takes into account the  abstentions excluded from the average. 

We next note that for any unitary transformation $U$ of the type defined after Eq.~(\ref{povm-pure}), the operators~$\{U\Pi_\chi U^\dagger,U\Pi_0 U^\dagger\}$ give the same value of~$Q$ and~$F(Q)$ as the original set~$\{\Pi_\chi,\Pi_0\}$, provided we change the guessing rule as $\vec n_\chi\to {\mathscr R}_U\vec n_\chi$, where~${\mathscr R}_U$ is the SO(3) rotation whose unitary representation is $U$. Therefore, one can easily prove that~$\Pi_0$ (the set $\{\Pi_\chi\}$) can always be chosen to be SU(2)  invariant  (covariant) by simply averaging over $U$.
In other words, with no loss of generality the POVM elements that provide a guess~$|\vec s\rangle$ can be chosen as
\begin{equation}
\tilde\Pi(\vec s)=U(\vec{s})\,\Pi\, U^\dagger(\vec{s}) ,
\end{equation}
where $\Pi\ge0$ is the so called seed of the POVM (in particular, note that~$\tilde\Pi(\hat z)=\Pi$). The abstention operator then reads
 \begin{equation}
\Pi_0=\openone- \int ds\, \tilde{\Pi}(\vec{s}) ,
 \end{equation}
which is manifestly  rotationally invariant (as claimed above). It is thus proportional to the identity on each
invariant subspace 
\begin{equation}
\Pi_0=\bigoplus_{j=j_{\rm min}}^J\; \bigoplus_{\alpha=1}^{n_j}a_{j\alpha} \openone_{j\alpha}=\bigoplus_{j=j_{\rm min}}^J a_j \openone_j,
\end{equation}
where $a_j$ are coefficients that satisfy the condition $0\leq a_j\leq 1$ and $\openone_j$ is defined in~Eq.~\eqref{completeness}. Here we have used the permutation invariance of the input state to fix, without loss of generality,~$a_{j \alpha}=a_{j}$ for all $\alpha$. We can also choose~$\tilde{\Pi}(\vec s)$ to have the block-diagonal form of the input state~$\tau(\vec n)$, namely,
\begin{equation}
\tilde{\Pi}(\vec{s})=\bigoplus_{j=j_{\rm min}}^J\bigoplus_{\alpha=1}^{n_j} \tilde{\Pi}_{j\alpha}(\vec{s}).
\end{equation}
For given $\{a_j\}$, the optimality of $\Pi_{j\alpha}(\vec{s})$, defined in Eq.~\eqref{Pi-j-alpha}, clearly ensures that
\begin{equation}
\tilde{\Pi}_{j\alpha}(\vec{s})=(1-a_j) \Pi_{j\alpha}(\vec{s}),
\end{equation}
are also optimal for estimation with abstention.
Recalling that the label~$\alpha$ is unsubstantial, aside from the multiplicative factor $n_j$, 
we  have from~(\ref{def Q})  that the abstention probability is simply
\begin{equation}\label{pQ}
Q=\sum_{j=j_{\rm min}}^J p_j a_j ,
\end{equation}
where $p_j$ is given in Eq.~\eqref{p-j}. The coefficients $a_j$ can be understood as the probabilities  of abstention conditioned to the input state having total angular momentum~$j$, 
i.e.,~\mbox{$a_j=p(\mbox{abstention}| j)$}. Similarly, for a given $j$, the 
probability of producing an estimate, or accepting, is
$\bar{a}_j=1-a_j=p(\mbox{acceptance}| j)$.

{}From Eq.~(\ref{F(Q)}) we obtain
\begin{equation}\label{F-Q}
F(Q)=\frac{1}{2} \left( 1 +  \sum_{j=j_{\rm min}}^J p_j\tilde{\Delta}_j\right),
\end{equation}
where
\begin{equation}\label{delta-p-j}
\tilde{\Delta}_j=\frac{1-a_j}{1-Q}\Delta_j=\frac{\bar{a}_j}{\bar{Q}}\Delta_j,
\end{equation}
and the quantity $\Delta_j$ is given in Eqs.~\eqref{Delta-j} and \eqref{m-bar}. Thus, we are only left with the free parameters $a_j$, which have to be optimized in order  to maximize $F(Q)$, subject to the constraints $0\le a_j\le 1$ and~(\ref{pQ}).
Somehow expected, one can show that~$\Delta_j$ is a monotonically increasing function of~$j$, i.e., $\Delta_{j-1}<\Delta_j$, therefore the largest contribution to the fidelity is given by $\Delta_{J}$. 
This corresponds to $\bar a_{J}=1$ and~$\bar a_j=0$, $j<J$. Hence, for unrestricted probability of abstention, 
the optimal protocol discards any contribution with $j< J$. This protocol, however, would 
provide an estimate with a  probability  that  decreases exponentially with $N$, for $r<1$, as $p_{J}\simeq (1/r)[(1+r)/2 ]^{N+1}$. 
Notice that in a noiseless scenario, $r=1$, there is only the contribution~$j=J$, which is already the optimal one and therefore abstention is of no use in
such case.

Clearly, for finite $Q$  there can be  contributions from other total angular momentum eigenspaces ($j<J$) compatible with Eq.~(\ref{pQ}).  Recalling the monotonicity of $\Delta_j$, and by convexity, it is obvious from Eqs.~(\ref{F-Q}) and~(\ref{delta-p-j}) that  there must exist an angular momentum threshold~$j^*$ such that~$\bar a_j=0$ ($\bar a_j=1$), if $j<j^*$ ($j>j^*$). 
The value $j^*$ is determined through Eq. \eqref{pQ}  to~be
\begin{equation}\label{j-star}
j^*=\max\left\{j  \  \mbox{such that}  \  Q-\kern-.1em\mbox{$\sum_{j'=j_{\mathrm{min}}}^{j-1} p_{j'}$} \geq 0 \right\}.
\end{equation}
Thus, we have
\begin{equation}
a_j=\left\{\begin{array}{ll}
1,& j<j^*; \\[.2em]
p_j^{-1}  \left(Q-\sum_{j'=j_{\mathrm{min}}}^{j^*-1}p_{j'}\right),& j=j^*; \\[.5em]
   0 ,& j>j^* .
   \end{array}
   \label{a-j}
   \right.
   \end{equation}
In a more physical language, the optimal strategy consists actually of two successive measurements. The experimentalist first measures the total angular momentum~$j$ of the input state $\tau(\vec n)$ and decides to abstain (provide a guess) if $j<j^*$ ($j>j^*$). If~$j=j^*$, she simply decides randomly, by tossing a Bernoulli coin with probability~$a_{j^*}$ of coming up heads, and if  heads (tails) show up, abstain (provide a guess). In order to provide the actual guess, if she decides to do so, she performs the optimal POVM measurement~$\{\Pi(\vec s)\}$ [or just $\{\bigoplus_\alpha\Pi_{j\alpha}(\vec s)\}$] in Eq.~(\ref{repeated}) on the state $\bigoplus_\alpha\tau_{j\alpha}(\vec n)$ that resulted from the first measurement.

\subsection{Small number of copies}\label{ss-small}

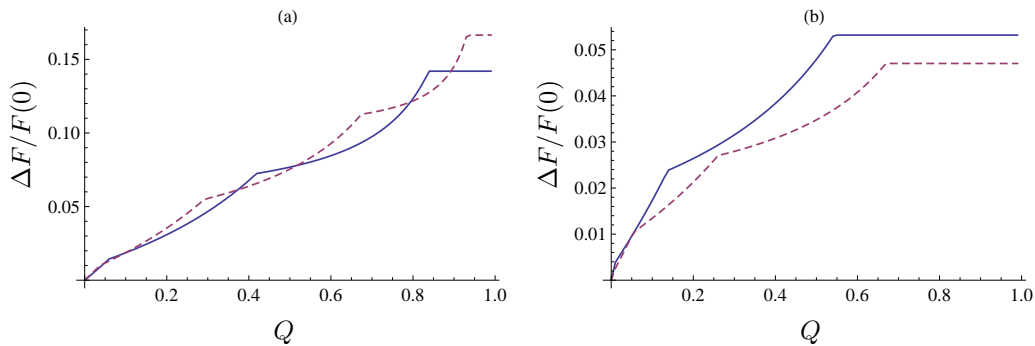
\begin{figure}
	\centering
	\setlength{\unitlength}{5mm}
\thinlines
\begin{picture}(25.,10)(0,0)
\put (0,1.0){\includegraphics[scale=.72]{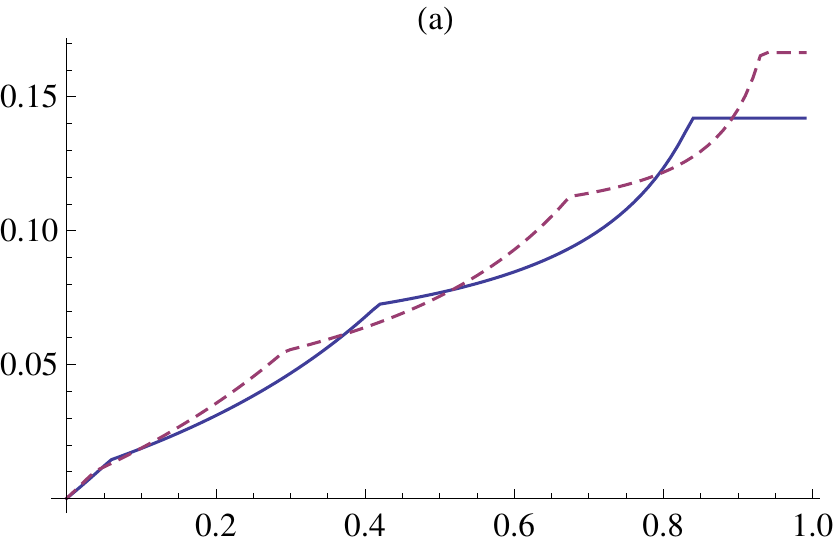}}
\put (14,1.){\includegraphics[scale=.72]{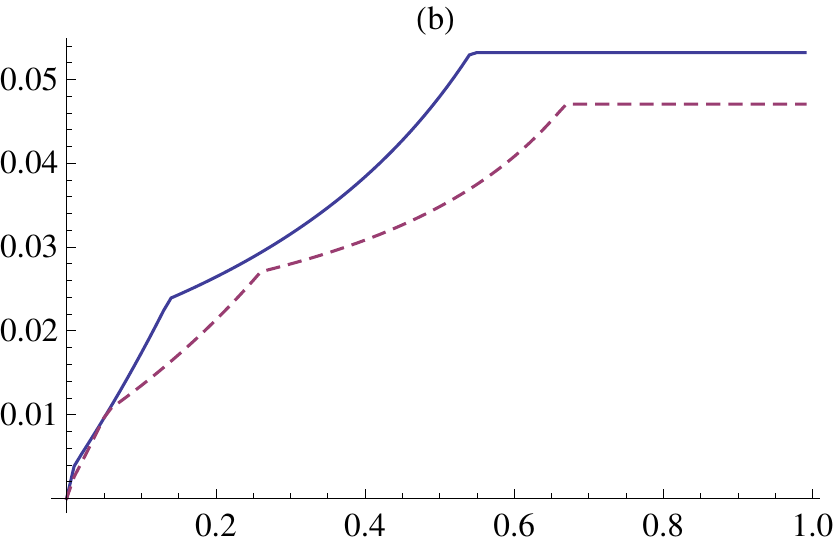}}
%
\put (-1.0,4.1){\rotatebox{90}{\footnotesize{$\Delta F/F(0)$}}}
\put (6.,.1){{\footnotesize $Q$}}
\put (13.,4.1){\rotatebox{90}{\footnotesize{$\Delta F/F(0)$}}}
\put (20.,.1){{\footnotesize $Q$}}
\end{picture}
\caption{\label{q-gain-03-07} Fidelity gain $\Delta F/F(0)=[F(Q)-F(0)]/F(0)$ as a function of $Q$ for
$N=6$ (solid line) and $N=8$ (dashed line) and purities of $r=0.3$ in (a) and  $r=0.7$ in (b).
} 
\end{figure}
\begin{figure}[b]
	\centering
	\setlength{\unitlength}{5mm}
\thinlines
\begin{picture}(15.5,11.2)(0,0)
\put (.8,.7){\includegraphics[scale=1.]{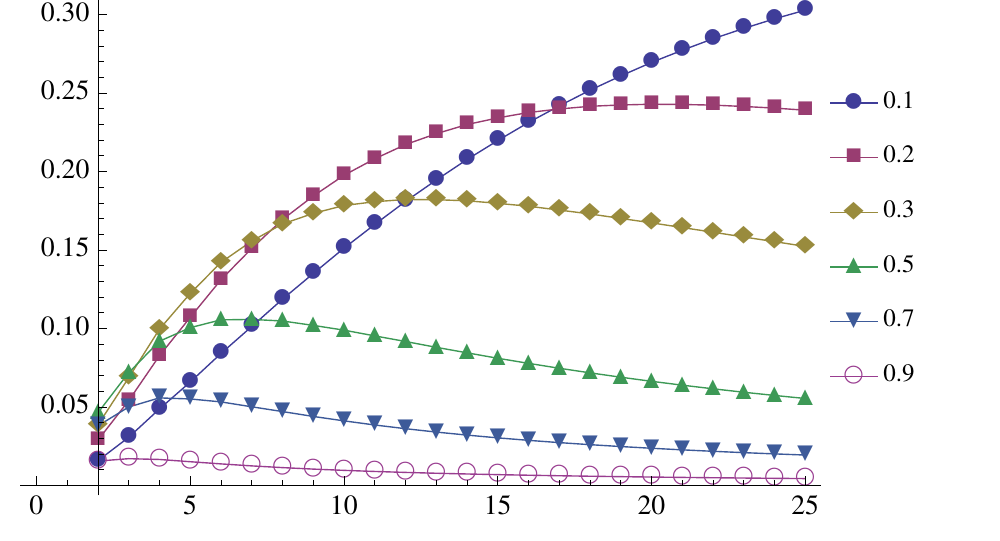}}
\put (-.4,6.){\rotatebox{90}{$\Delta F/F(0)$}}
\put (9.8,.3){$N$}
\end{picture}
\caption{\label{f-gain} Fidelity gain as a function of $N$ for various values of $r$, indicated in the legend, and for~$Q\ge Q_{\rm crit}$. }
\end{figure}

In Fig.~\ref{q-gain-03-07} we plot the fidelity gain due to abstention  $[F(Q)-F(0)]/F(0)=\Delta F/F(0)$ vs.~$Q$ for $N=6$, $N=8$, and purities of $r=0.3$ and $r=0.7$.
The structure of~Eq.~(\ref{a-j}) is  apparent from these plots:  at $Q=0$ ($a_j=0$ for all $j$) there is, naturally, no gain;
kinks sequentially appear at the precise values of $Q$  where a new coefficient $a_j$ in~(\ref{a-j}) 
becomes positive (and $j^*$ increases by one); the curves are convex between successive kinks, where the one $a_j$ that has become positive, $a_{j^*}$,  keeps increasing.
This pattern repeats until the abstention rate $Q$ reaches a critical value~$Q_{\mathrm{crit}}$ at which $j^*=J$, 
 \begin{equation}\label{q-crit}
 Q_{\mathrm{crit}}=1-p_{J}=1-{1\over r}\left(\frac{1+r}{2}\right)^{2J+1}\kern-1em+{1\over r}\left(\frac{1-r}{2}\right)^{2J+1}
 \end{equation}
[see Eq.~\eqref{p-j}]. Ê
Increasing $Q$ further will not provide any additional gain,  as the flat
plateaus of  Fig.~\ref{q-gain-03-07} illustrate. 
This is so, since one can view the optimal abstention protocol as a filtering process where the low angular momentum components of the input state are filtered out. Hence,  keeping the maximum value of~$j=J$ is the optimal filtering beyond which no further improvement is possible.
Fig.~\ref{q-gain-03-07}(a)  shows that in noisy scenearios, e.g.
 $r = 0.3$, abstention can increase the fidelity quite notably, up to 15\%. 
For higher purities the gain is more moderate, as shown in  Fig.~\ref{q-gain-03-07}(b). The enhancement in this case is about 4-5\% but with an abstention rate slightly above~50\%.
Further results are shown in Fig.~\ref{f-gain}, where we plot the fidelity gain as a function of the  number of copies $N$ for abstention rates larger than $Q_{\mathrm{crit}}$, and for various values of the purity $r$. All the curves have a maximum at a value of $N$ that  varies with the purity. The lower the purity, the higher the value of $N$ at which the maximum occurs (e.g., for $r=0.3$ the maximum gain occurs at $N=12$; for $r=0.1$ the maximum is off scale at the right of the figure).

As we have seen, the possibility of abstaining enables us to reach values of the fidelity that otherwise we could only attain with lower levels of noise. To quantify this effective reduction of noise, let us define an effective purity  $r_{\mathrm{eff}}$ by the implicit equation $F(r_{\rm eff},N,0)=F(r,N,Q)$.
That is, for an estimation setting, given by~$r$,~$N$,and~$Q$, $r_{\mathrm{eff}}$ is the purity of the input states 
that would provide the same fidelity if the standard strategy without abstention ($Q = 0$) were used instead.  Since $r$ is related to the probability of error $\eta$  in our model of noisy measurements in~\eqref{noise}, an increase of the effective purity corresponds to an effective reduction of the amount of  noise in the measurement through the relation $\eta_{\mathrm{eff}}=(3/4) (1-r_{\mathrm{eff}})$.
Figure~\ref{fig:reff} shows a plot of the effective purity $r_{\rm eff}$ as a function of $Q$ for various values 
of~$r$ and~$N$. 
As can be seen,~$r_{\mathrm{eff}}$ increases faster at low values of~$N$, but it saturates earlier (lower $Q_{\mathrm{crit}}$), reaching a lower value. For low $N$ and for a wide range of purities, $0.1\lesssim r\lesssim0.9$, we observe a constant effective increase of the purity, $r_{\mathrm{eff}}\approx r+0.2$, for reasonable values of the abstention rate~$Q$. As~$N$ increases one has to go to higher values of the abstention rate, $Q\sim Q_{\mathrm{crit}}$, to have a significant gain.  Hence, a moderate abstention rate is  most effective in noisy scenarios when a small, but fair, number of copies is available. 
 
Finally, let us point out that the protocol we have presented requires a projection on the total angular momentum eigenspaces.
This is a non-local measurement that nonetheless can be implemented efficiently~\cite{bacon}.
In a more extreme scenario where there are no restriction on the abstention rate, one can  attain the maximum fidelity with an even simpler strategy:
perform a local Stern-Gerlach measurement on every qubit (say, of the $z$-component of the spin)  and abstain unless all outcomes agree. This strategy renders an abstention probability of $Q=1-[(1+r)/2]^{N}$, which might be comparable to~$Q_{\mathrm{crit}}$ in~Eq.~\eqref{q-crit}.

\begin{figure}
	\centering
	\setlength{\unitlength}{5mm}
\thinlines
\begin{picture}(19,20)(0,0)
\put (3.2,0.6){\includegraphics[scale=1.]{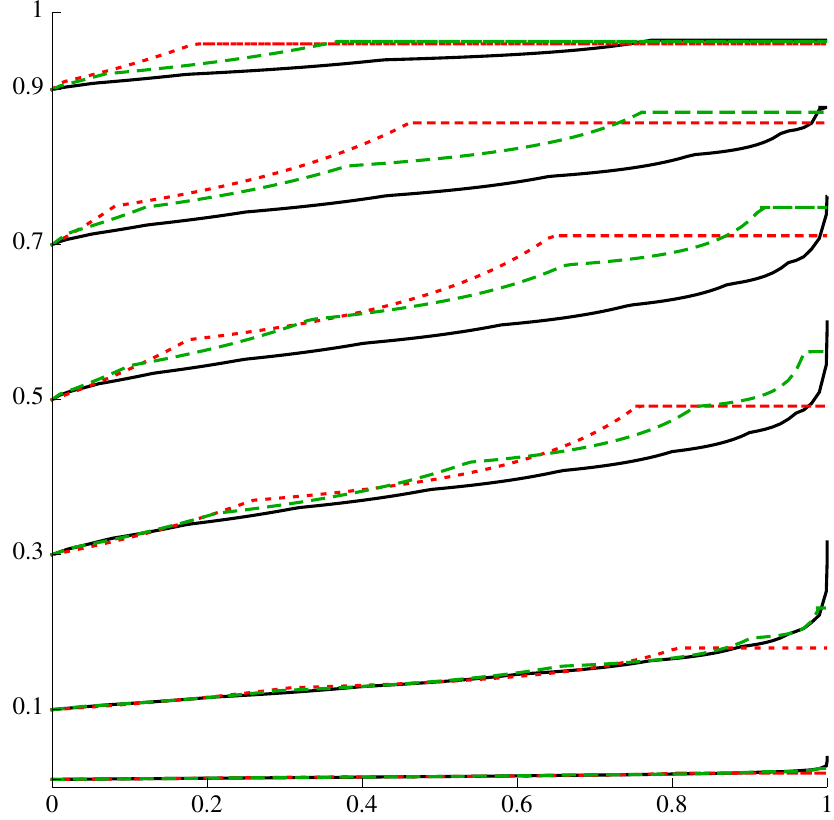}}
\put (2,9.9){\rotatebox{90}{{\large $r_{\rm eff}$}}}

\put (12.5,.0){$Q$}
\end{picture}
\caption{\label{fig:reff} 
Effective purity $r_{\mathrm{eff}}$ as a function of the abstention
rate~$Q$ for $N = 5$ (dotted), $10$ (dashed), and Ê$30$ (solid), and for
purities of $r = 0.01,\ 0.1,\ 0.3,\ 0.5,\ 0.7$, and~$0.9$, which can be read off from the values of $r_{\mathrm{eff}}$ at~$Q=0$.
} 
\end{figure}
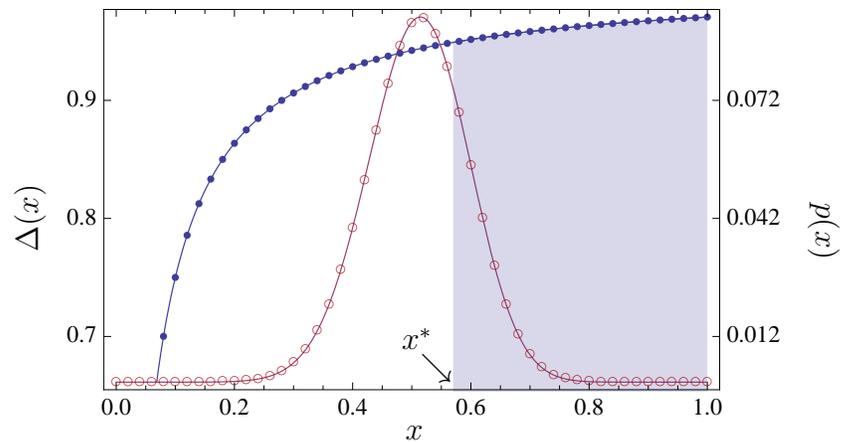
\begin{figure}
	\centering
	\setlength{\unitlength}{5mm}
\thinlines
\begin{picture}(17.3,13)(0,0)
\put (2,.5){\includegraphics[scale=1.12]{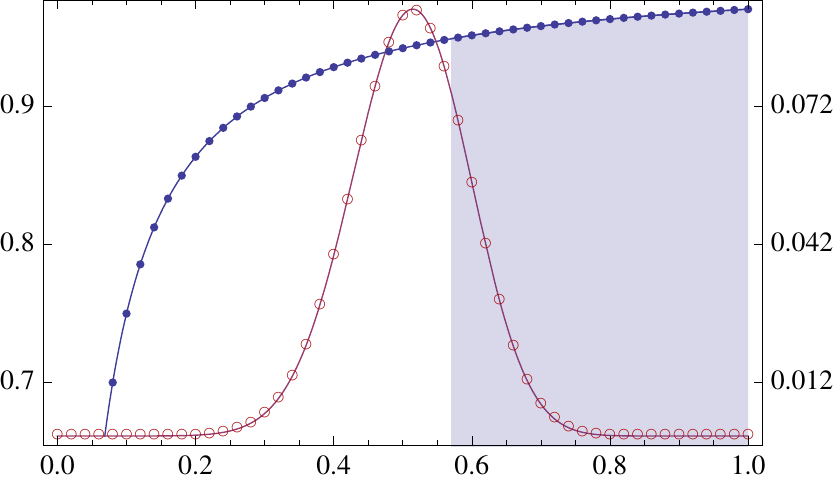}}
\put (11.4,1.7){$\searrow$}
\put (10.9,2.4){$x^*$}
\put (.6,5.){\rotatebox{90}{$\Delta(x)$}}
\put (21.7,6.3){\rotatebox{270}{$p(x)$}} 
\put (11,.0){$x$}
\end{picture}
\caption{\label{fal}
Plots of  $\Delta(x)$ (blue line with solid circles) and $p(x)$ (red line with empty circles) for $N = 100$ and $r = 0.5$. The circles represent the quantities $\Delta_j$ and $J p_j$ as a function of $x= j/J$. The shaded area indicates the acceptance region for an abstention rate $Q\sim 93\%$.
} 
\end{figure}

\subsection{Asymptotic regime}

We next compute the analytical expressions of the fidelity in the large $N$ limit.
Here it is useful to define the variable $x$ as
\begin{equation}\label{x}
x={j\over J},\quad 0\le x\le 1,
\end{equation}
which becomes continuous in the limit~$N\to\infty$ ($J\to\infty$). In this case, we can replace~$p_j$  by the  continuous probability distribution in $[0,1]$ defined by
 \begin{equation}\label{p-x}
  p(x)=J p_{j=x J},
  \end{equation}
so that $\int_0^1 dx\,p(x)=1$ as $N$ goes to infinity.  Eq.~(\ref{F-Q}) can then be approximated by its continuous version, which reads
 \begin{equation}\label{F(Q) cont}
 F={1\over2}\left[1+\int_0^1 dx\, p(x)\tilde\Delta(x)\right],
 \end{equation}
 where
 \begin{equation}
 \tilde\Delta(x)=\tilde\Delta_{j=x J},
 \label{tD(x)}
 \end{equation}
 where recall that $\tilde\Delta_j$ is given in Eqs.~\eqref{delta-p-j}.
{}From Eq.~(\ref{a-j}) we see that asymptotically $\bar a_j$ becomes the step function $\theta(x-x^*)$, where
$x^*=j^*/J$, and we have used the standard definition
\begin{equation}
\theta(x)=\left\{
\begin{array}{ll}
1,&x\ge0;\\
0,&x<0 .
\end{array}
\right.
\end{equation}
With this, Eq.~(\ref{delta-p-j}) becomes
\begin{equation}
\tilde\Delta(x)={\theta(x-x^*)\over \bar Q}\Delta(x), 
\end{equation}
 and,  in turn,
 \begin{equation}\label{Fx*}
 F={1\over2}\left[1+{1\over\bar Q}\int_{x^*}^1dx\,p(x)\Delta(x)\right].
 \end{equation}
 It also follows from~(\ref{pQ}) that
 \begin{equation}\label{barQ theta}
 \bar Q=\int_0^1 dx\, p(x)\theta(x-x^*)=\int_{x^*}^1 dx\, p(x) .
 \end{equation}

At this point, we need to find a good approximation to $p(x)$ that would enable us to obtain the explicit form for the asymptotic fidelity.
{}From Eq.~\eqref{p-j}, and using the Stirling formula, we obtain 
  \begin{equation}\label{p-x-2}
 p(x)\simeq \sqrt{\frac{N}{2 \pi}}\frac{1}{\sqrt{1-x^2}} {x(1+r)\over r(1+x)}\;\mathrm{e}^{- N H(\frac{1+x}{2}\parallel\frac{1+r}{2})},
 \end{equation}
 where $H(s\parallel t)$ is the (binary) relative entropy
 \begin{equation}
H(s\parallel t)=s \log\frac{s}{t}+ (1-s)\log\frac{1-s}{1-t} ,
\end{equation}
and the approximation  is valid for both $x$ and $r$ in the open unit interval~$(0,1)$.
The appearance of a relative entropy in Eq.\eqref{p-x-2} can be understood as follows. Our $N$-copy input state (diagonal in the canonical $J_{n}$ basis) can be thought of as a classical coin tossing distribution of $N$ identical coins with a bias of~$(1+r)/2$. From the theory of types~\cite{cover} it is well known that the probability to get~$k$ heads is given by the Kulback-Leibler distance (or relative entropy) between the empirical distribution $\{f=k/N, 1-f\}$  and the distribution  $\{(1+r)/2,(1-r)/2\}$. That is,  $p(k)\sim \exp \{-N H[ f \parallel (1+r)/2]\} $ to first order in the exponent.
The number of heads~$k$ is in one-to-one correspondence with the magnetic quantum number, $m=k-J$, and  the conditioned probability~$p(j|m)$ is strongly peaked at $m=j$, as one can easily check. It follows that the probability that the input state has total angular momentum $j$, given by~$p(j)=\sum_{m} p(j|m)p(m)$, will be asymptotically determined by the probability distribution~$p(m)$, which has a convenient expression in terms of the typical and the empirical distribution of up/down outcomes.

{}From Eq.~\eqref{p-x-2} it follows that $p(x)$ is peaked at the value~$x=r$, i.e. at $j=r J$, as shown in Fig.~\ref{fal} and stated without a proof in Sec.~\ref{sec:no-abstention}. Actually, around the peak,~$x\sim r$, the exponent becomes quadratic and~$p(x)$ approaches the 
Gaussian distribution
\begin{equation}\label{pGauss}
p(x)\simeq\sqrt{N\over2\pi(1-r^2)} \mathrm{e}^{ -N \frac{(x-r)^2}{2(1-r^2)}},
\end{equation}
 as also follows from the central limit theorem, whereas it falls off exponentially elsewhere.

It is now apparent that, asymptotically, abstention has negligible impact Êif components with $j$ below $rJ$ are filtered out ($x^*<r$), since Êthe main contribution to the fidelity, which comes from the peak around~$x\simeq r$, is not excluded from the integral in~Eq.~(\ref{Fx*}) (only the left exponentially decaying tail is). 
For the same reason [see Eq.~(\ref{barQ theta})], $\bar Q\simeq 1$ (the abstention rate $Q$ is exponentially small), and~Eq.~(\ref{Fx*}) yields 
\begin{equation}\label{F(Q)2a}
F = 1- \frac{1}{2N}\frac{r+1}{r^{2}}+\ldots \quadÊ\mbox{ for } \; x^*< r,
\end{equation}
which is the same expression as the asymptotic fidelity of the protocol without abstention, Eq.~(\ref{fid-p}).

It is then clear that, in order to have a discernible improvement in the fidelity, the abstention threshold $x^*$ 
must lie to the right of the peak of the probability distribution. The fidelity in~(\ref{Fx*}) then can be written as
\begin{equation}\label{F(Q)2}
FÊ Ê \simeq Ê\frac{1}{2} \left[1+ {p(x^{*})\over\bar Q}\Delta(x^{*})\right] \simeq \frac{1}{2}\left[1+\Delta(x^{*})\ \right],\quad x^*> r, 
\end{equation}
where we have used that for $x\ge x^*>r$ and for large enough $N$, $p(x)$ falls off exponentially and the integral can be approximated by the value of the integrand at its lower limit. By the very same argument Eq.~(\ref{barQ theta}) gives
\begin{equation}
\bar Q\simeq Êp(x^*),
\end{equation}
which has also been used in~(\ref{F(Q)2}). Using now~(\ref{p-x-2}) we obtain that in the asymptotic limit of many copies, the rate at which our protocol provides a guess is
 \begin{equation}\label{qbar}
 \bar{Q}\sim \exp\left[- N H\left(\mbox{$\frac{1+x^*}{2}\parallel \frac{1+r}{2}$}\right)\right]  .
 \end{equation}
Recalling Eqs. \eqref{Delta-j} and \eqref{m-bar} we obtain the optimal fidelity:
\begin{equation}\label{fid-ass}
 F=1- \frac{1}{2Nx^{*}}\frac{r+1}{r}+\cdots ,\quad Ê\mbox{for $r\leq x^{*}\leq 1$},
\end{equation}
for a value of $Q$ given by \eqref{qbar}.
For $x^{*}=r$ the results \eqref{fid-p} Êand~(\ref{F(Q)2a}) are recovered, whereas for $x^{*}\to 1$ ($Q\geq Q_{\mathrm{crit}}$) the maximum average fidelity is attained
 \begin{equation}\label{f-max}
 F_{\mathrm{max}}=1- \frac{1}{2N }\frac{r+1}{r}+\cdots .
 \end{equation}

The advantage provided by our estimation with abstention protocol can be quantified by the effective number of copies that the standard protocol without abstention would require to achieve the same fidelity: Ê$ N_\mathrm{eff}=(x^*\kern-.2em/r) N$, where~\mbox{$x^*\in[r,1)$} is determined by the abstention rate $Q$ through~\eqref{qbar}. For high noise levels (low \mbox{purity,~$r\ll 1$}) our protocol provides an important saving of resources/copies, as~\mbox{$N_{\mathrm{eff}}/N=1/r \gg 1$}, whereas for nearly ideal detectors the saving in this asymptotic regime is more modest.

Alternatively, the advantage discussed above can also be quantified by the effective measurement-noise reduction, or equivalently, the effective purity~$r_{\rm eff}$ (See Sec.~\ref{ss-small}). Using \eqref{fid-ass} one can easily find a simple expression for the effective purity in the asymptotic limit and for large abstention rate: $r_{\mathrm{eff}} = (r+\sqrt{4r+5 r^{2}})/[2 (1+r)]$. In the limit of very low noise levels the errors probability $\eta$  [recall Eq.~\eqref{noise}] is effectively reduced by a factor of three, i.e., $\eta_{\rm eff}=\eta/3$, while in the opposite limit of very noisy measurements one finds $r_{\mathrm{eff}}=\sqrt{r}$. 

\subsection{Other regimes}
 
In the previous section we have seen how a gain in fidelity can be obtained  provided the `acceptance' rate $\bar Q$ falls off exponentially as $N$ becomes very large. Here we give an example where this gain takes place even at finite~$\bar Q$.

At fixed noise level (purity $r$), the fidelity is an increasing function of $N$. However, one could imagine an experimental setup where the noise (purity) also increases  (decreases) with $N$. If this is so,  the asymptotic fidelity could be strictly less than one, or in other words, perfect estimation could be unattainable  even with unbounded resources. This is the case in our example, were we assume that  $r={a}/\sqrt N$, $a$ being a positive constant.  
Notice that the  threshold $x^*$ must also scale as $1/\sqrt{N}$ in order to have a reasonably low abstention rate. Therefore, it is convenient to use 
a new variable~\mbox{$\xi=\sqrt{N} x=\sqrt N\, j/J=2j/\sqrt N$} instead. Then, the probability distribution in  this new variable  is 
\begin{equation}
p(\xi)={\sqrt N\over2} p_{j=\xi\sqrt N/2},\quad\mbox{with $\displaystyle r={a\over\sqrt N}$}. 
\end{equation}
Recalling 
Eq.~\eqref{p-j}  and using Stirling formula this equation gives
\begin{equation}
p(\xi)={{\rm e}^{-\left({\xi-{a}\over\sqrt2}\right)^2}-{\rm e}^{-\left({\xi+{a}\over\sqrt2}\right)^2}\over\sqrt{2\pi}{a}}\xi
\end{equation}
to leading order in inverse powers of $N$. The subleading terms are of order $N^{-1/2}$ and will be neglected here.
For a given threshold value $\xi^* =2 j^*/\sqrt{N}$  the abstention rate is
\begin{equation}\label{q-xi}
Q=\int_0^{\xi^*}p(\xi)\,d\xi={1\over2}\left({\rm erf}\,{\xi^*_+}+{\rm erf}\,{\xi_{-}^*}\right)-{{\rm e}^{-{\xi^*_-}^2}-{\rm e}^{-{\xi^*_+}^2}\over\sqrt{2\pi}{a}} ,
\end{equation}
where 
${\xi^*_{\pm}}=(\xi^*\pm{a})/\sqrt2$ and $\mathrm{erf}\, x$ is the error  function.

{}From Eqs.~\eqref{Delta-j} and \eqref{m-bar} we have in this same regime and at leading order
\begin{equation}
\Delta(\xi)=\Delta_{j=\xi\sqrt N/2}=1-{2\over1-{\rm e}^{2{a}\xi}}-{1\over{a}\xi} .
\end{equation}
With the above, the fidelity~(\ref{F-Q}), [or rather, the counterpart of~(\ref{F(Q) cont})] 
is
\begin{equation}
 F={1\over2}\left[1+\int_0^\infty d\xi\, p(\xi)\tilde\Delta(\xi)\right]=
 {1\over2}\left[1+{1\over\bar Q}\int_{\xi^*}^\infty d\xi\, p(\xi)\Delta(\xi)\right],
\end{equation}
where the last integral can be computed to be
\begin{eqnarray}\label{mean-delta-xi}
\Delta^*&\equiv&\int_{\xi^*}^\infty\Delta(\xi)p(\xi)\,d\xi=\\
&=&{1-{a}^2\over2{a}^2}\left({\rm erf}\,{\xi^*_-}-{\rm erf}\,{\xi^*_+}\right)
+{{\rm e}^{-{\xi^*_-}^2}+{\rm e}^{-{\xi^*_+}^2}\over\sqrt{2\pi}{a}} .\nonumber
\end{eqnarray}
\begin{figure}[t b]
	\centering
	\setlength{\unitlength}{5mm}
\thinlines
\begin{picture}(16,10)(0,0)
\put (1,.5){\includegraphics[scale=.85]{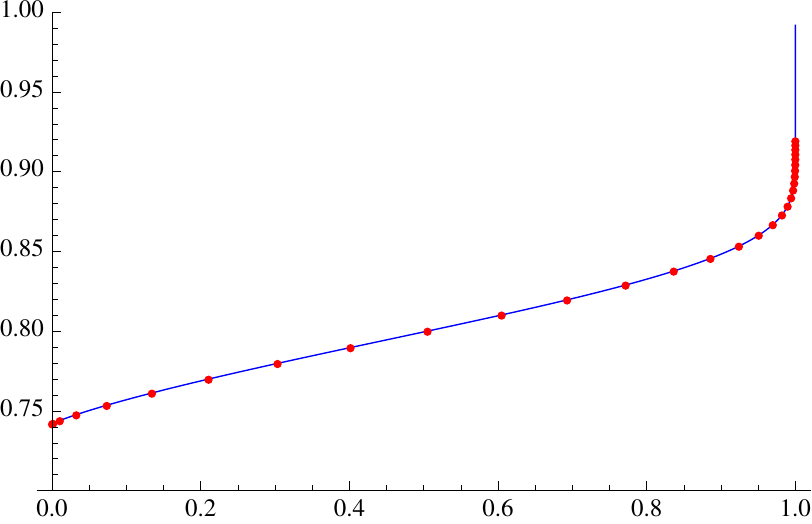}}
\put (.1,9.5){$F$}
\put (15.5,.2){$Q$}
\end{picture}
\caption{\label{weird} Plot of the fidelity as a function of $Q$ for $r=a/\sqrt N$, with the choice~${a}=1.0$, $N=10^{6}$ (red circles). The solid line (in blue) is the leading term in Eq.~\eqref{fid-weird} plotted  as a function of $Q$ [a parametric plot of the pairs $(Q,F)$, as given by Eqs.~(\ref{q-xi}) and~(\ref{fid-weird})]
} 
\end{figure}
We can finally write the fidelity as 
\begin{equation}\label{fid-weird}
F={1\over2}\left(1+\frac{\Delta^*}{1- Q}\right)+{\mathscr O}(N^{-1/2}) .
\end{equation}
As shown in~(\ref{q-xi}) and~(\ref{mean-delta-xi}), both $Q$ and $\Delta^*$ are functions of the filtering  threshold~$\xi^*$, which is just a properly scaled version of~the original threshold~$j^*$. Finding the maximum fidelity for a given rate of abstention $Q$ requires inverting Eq.~(\ref{q-xi}) to obtain~$\xi^*\!(Q)$, but this cannot be done analytically and one has to resort to  numerical methods.

In Fig.~\ref{weird} we plot $F$ as a function of $Q$ for $a=1$.
The increase of the fidelity in the asymptotic regime of large $N$ is clearly seen: e.g., an abstention rate of a 50\%  yields  a
rise of about $10\%$, and it goes up to about 30\% for higher (but still reasonable) values of~$Q$.
The figure also shows 
the agreement between the approximate form of the fidelity given by Eqs.~\eqref{q-xi} to~\eqref{fid-weird}  and the numerical evaluation of its exact expression in~(\ref{F-Q}).

It should be noted that in the regime described here a rise of the input size $N$ fails to replicate the fidelity improvement that results from increasing the rate of abstention (no $N_{\rm eff}$ can be defined in this regime), thus abstention appears to be the only means by which one can improve estimation. 

\section{Conclusions}

In this work we have addressed optimal estimation of pure qubit states when abstention from providing an outcome is allowed. We have considered a reasonably realistic multiple-copy scenario, where a sample of $N$ identically prepared systems go through a  non-ideal (noisy) process of measurement. 
We have  shown that in the limit of zero noise, abstention does not help to improve estimation (it does not hamper it either). However, abstention turns out to counterbalance the adverse effect of errors in a noisy process of measurement.
We  have shown that in general abstention is most useful for inputs of few copies and for error rates of the order of  a few percent.
E.g., for~$N=6$ and a value of the error probability of $\eta=0.5$  (per qubit),  one can easily attain fidelity gains of the order of~$15\%$ with an abstention rate of~$Q=4/5$. As $N$ increases, one needs to allow for higher abstention rates to obtain a significant improvement. We have given analytical asymptotic expressions of the fidelity valid in the limit of large number of copies. 
In this limit, abstention can have the effect of increasing  the number of copies by a constant fraction: $N_{\mathrm{eff}}/N=x^{*}/r$ ($x^{*}>r$), with an acceptance rate~$\bar Q$ given by the relative entropy: $-(1/N)\log\bar{Q}= H[(1+x^{*})/2\parallel (1+r)/2]$. For low levels of noise this amounts to reducing the error probability $\eta$ by a factor of up to three.

We have also considered a scenario where the noise (per qubit) increases with the number of copies  in such a way that perfect estimation is unattainable \mbox{($\lim_{N\to\infty}F<1$).} In this case one can obtain a significant enhancement of the asymptotic fidelity (few percent) even for finite abstention probabilities $Q<1$. Moreover, in such scenario abstention appears to be the only way to improve estimation.

In broader parameter estimation contexts, where, e.g., Êphase or direction information is encoded in more general many-particle states~\cite{bagan-optimal}, Êabstention may have a much more dramatic effect. These issues will be analyzed in a separate publication~\cite{in-preparation-1}.

\ack
We acknowledge financial support from ERDF: European Regional Development Fund. This research was supported by
 the Spanish MICINN, through contract FIS2008-01236 and the Generalitat de
Catalunya CIRIT, contract  2009SGR-0985. We thank G. Chiribella for useful discussions.

\end{document}